\documentclass[
preprint,
showpacs,preprintnumbers,
bibnotes,
 amsmath,amssymb,
 aps,
 pra,
superscriptaddress,
longbibliography,
]{revtex4-1}

\usepackage{amsmath}
\usepackage{amsfonts}
\usepackage{amssymb}
\usepackage{graphicx}
\usepackage{xspace}
\usepackage{color}
\usepackage{comment}
\usepackage[hidelinks]{hyperref}
\usepackage{dcolumn}
\usepackage{bm}

\usepackage{comment}

\usepackage[detect-weight=true, detect-family=true]{siunitx}

\begin{document}

\title{
Hybrid tip-enhanced nano-spectroscopy and -imaging of monolayer WSe$_2$ with local strain control
}

\author{Kyoung-Duck Park}
\affiliation
{Department of Physics, Department of Chemistry, and JILA,\\ University of Colorado, Boulder, CO, 80309, USA}
\author{Omar Khatib}
\affiliation
{Department of Physics, Department of Chemistry, and JILA,\\ University of Colorado, Boulder, CO, 80309, USA}
\author{Vasily Kravtsov}
\affiliation
{Department of Physics, Department of Chemistry, and JILA,\\ University of Colorado, Boulder, CO, 80309, USA}
\author{Genevieve Clark}
\affiliation
{Department of Physics, Department of Materials Science and Engineering, University of Washington, Seattle, Washington 98195, USA}
\author{Xiaodong Xu}
\affiliation
{Department of Physics, Department of Materials Science and Engineering, University of Washington, Seattle, Washington 98195, USA}
\author{Markus B. Raschke}
\affiliation
{Department of Physics, Department of Chemistry, and JILA,\\ University of Colorado, Boulder, CO, 80309, USA}
\email{markus.raschke@colorado.edu}
\date{\today}

\begin{abstract}
\noindent
Many classes of two-dimensional (2D) materials have emerged as potential platforms for novel electronic and optical devices. 
However, the physical properties are strongly influenced by nanoscale heterogeneities in the form of edges, grain boundaries, and nucleation sites.
Using combined tip-enhanced Raman scattering (TERS) and photoluminescence (TEPL) nano-spectroscopy and -imaging, we study the associated effects on the excitonic properties in monolayer WSe$\rm{_2}$ grown by physical vapor deposition (PVD).
With ${\sim}15$ nm spatial resolution we resolve nonlocal nanoscale correlations of PL spectral intensity and shifts with crystal edges and internal twin boundaries associated with the expected exciton diffusion length.
Through an active atomic force tip interaction we can control the crystal strain on the nanoscale, and tune the local bandgap in reversible (up to 24 meV shift) and irreversible (up to 48 meV shift) fashion.
This allows us to distinguish the effect of strain from the dominant influence of defects on the PL modification at the different structural heterogeneities. 
Hybrid nano-optical and nano-mechanical imaging and spectroscopy thus enables the systematic study of the coupling of structural and mechanical degrees of freedom to the nanoscale electronic and optical properties in layered 2D materials.

\end{abstract}

\maketitle

\section{Introduction}
\noindent
Layered two-dimensional (2D) transition metal dichalcogenides (TMDs) have emerged as a new platform for studying quantum confined semiconductor physics \cite{cui2015, moody2015, qian2014, kang2015, koppens2014}. 
As the TMD crystals are thinned to the monolayer (ML) limit, new properties emerge including an indirect-to-direct bandgap transition \cite{mak2010atomically, splendiani2010, wang2012}, valley-specific circular dichroism \cite{mak2012control, kuc2011, jones2013, xu2014spin}, or an enhanced nonlinear optical response \cite{yin2014, cheng2015}. 
The direct semiconducting gap, large spin-orbit coupling, and valley-selectivity provide several advantages for the use of TMDs in photodetector and other optoelectronic device applications. 

A prevailing theme in TMDs and other layered van der Waals systems is the complex interaction between fundamental excitations inherent to the materials themselves, and extrinsic factors associated with surface morphology and the underlying substrate. 
The reduced dimensionality invites strong interference from charged impurities, defects, and disorder, creating much difficulty in isolating the intrinsic quantum properties of the material system \cite{chow2015}. 
The resulting electronic properties are consequently highly inhomogeneous and sensitive to structural variations near internal and external boundaries \cite{van2013, liu2014}. 
To explore these heterogeneities and how they control the optical and electronic properties, a comprehensive multi-modal nano-scale imaging and spectroscopy approach is desired. 

High-resolution local probes such as scanning tunneling microscopy (STM) and transmission electron microscopy (TEM) uncover the specific nature of structural defects and grain boundaries (GBs) that may lead, for example, to an increase or decrease in the electrical conductivity \cite{van2013, najmaei2013, huang2015}. 
However, despite atomic-scale spatial resolution, these experimental techniques provide limited information of the associated electronic, spin, or optical response. 
To this end, a series of recent studies applying confocal \cite{van2013, liu2014, park2015} and near-field photoluminescence (PL) mapping \cite{bao2015, Lee2015}, with spatial resolution reaching as high as ${\sim}60$ nm, have addressed the question of the local modification of the optical and electronic properties at GBs.
However, the observations of both increases and decreases in PL yield for different experimental conditions have left a confusing picture regarding the relative role of doping, defects, mid-gap exciton states, or strain controlling the PL behavior at structural heterogeneities.  

Here we present a new hybrid nano-opto-mechanical tip-enhanced spectroscopy and imaging approach combining nano-Raman (tip-enhanced Raman scattering, TERS), nano-PL (tip-enhanced photoluminescence, TEPL), and atomic force local strain control to investigate the correlation of local structural heterogeneities with nanoscale optical properties with enhanced ${\sim}15$ nm spatial resolution. 
Using a novel tilted tip approach for in-plane near-field polarization control, we study the excited state PL response in twinned WSe$\rm{_2}$ ML physical vapor deposition (PVD) grown micro-crystals. 
A combination of PL quenching and selective spectral changes at crystal edges, twin boundaries, and nucleation sites (NS) is resolved on 10's nm length scales, associated with the theoretical exciton diffusion length (${\sim}24$ nm \cite{van2013}).
Through controlled tip-sample force interaction we can tune the bandgap reversibly (up to 24 meV) and irreversibly (up to 48 meV) through local nanoscale strain engineering (0 - 1\%). 
The combined results allow for the separation of the effect of strain from controlling the PL modification at edges, NS, and internal twin boundaries, and suggest defects and stoichiometry as the primary factors modifying the PL at the structural heterogeneities, yet in distinctly different ways. 
The application to WSe$_2$ shows the potential of combined nano-optical and nano-mechanical spectroscopy and imaging with nanometer spatial resolution, few cm$^{-1}$ spectral resolution, and nN force sensitivity with a wide application range extending beyond 2D materials.

\section{Experiment}
\noindent
\begin{figure}
	\includegraphics[width = 13cm]{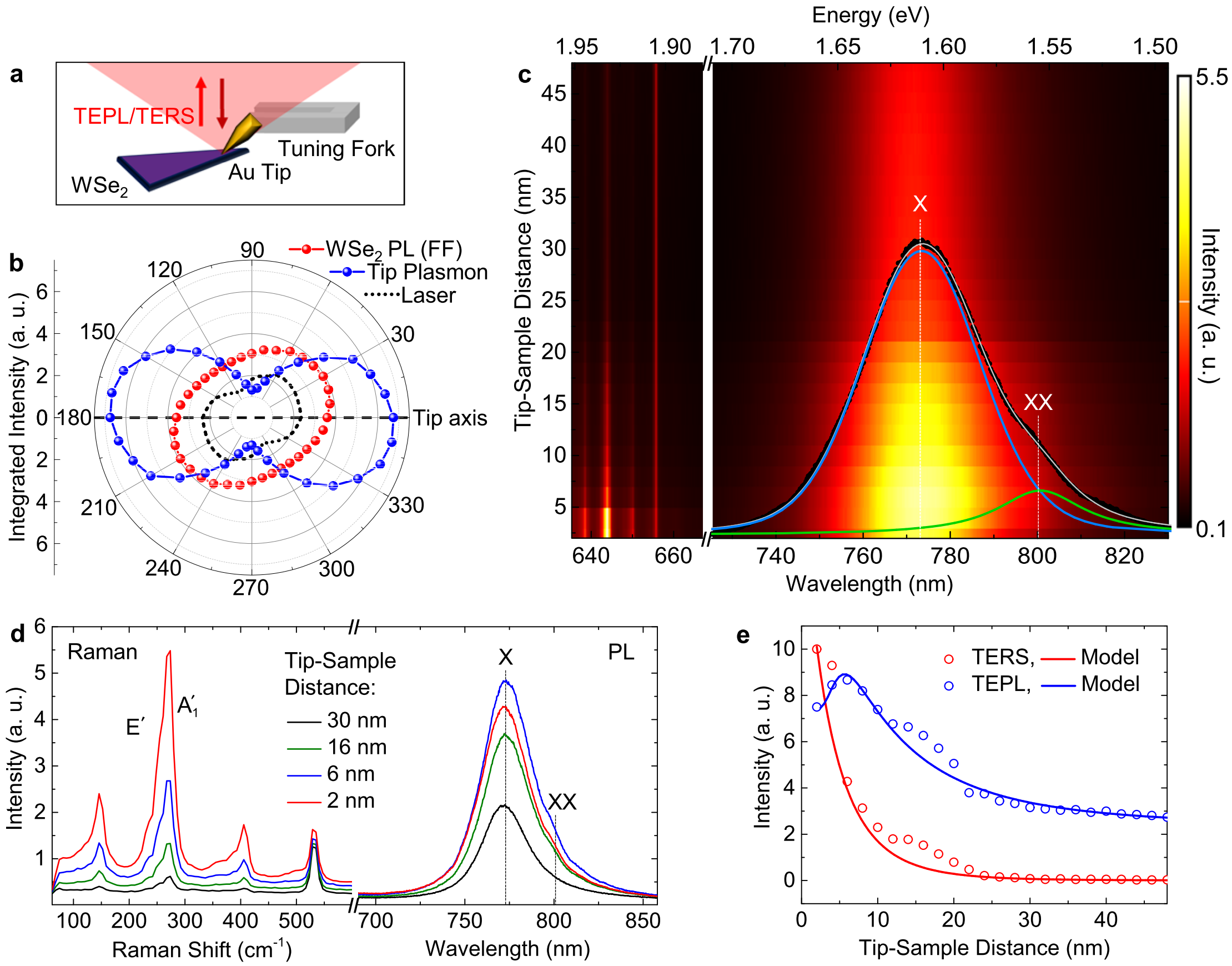}
	\caption{
Schematic of multi-modal TEPL/TERS (a) with polar plot of the integrated intensity (b) for the tip plasmon (blue line), WSe$\rm{_2}$ PL (red line), and excitation laser (black line).
(c) Tip-sample distance dependence of TEPL and TERS of monolayer WSe$\rm{_2}$. 
Overlapped TEPL spectrum (black) at 6 mm distance with Voigt profile fit (gray) decomposed into exciton (blue, X), and possibly biexciton (green, XX) emission.
(d) Selected TERS and TEPL spectra for different tip-sample distances. 
(e) Peak intensity dependence of WSe$\rm{_2}$ Raman (273~\si{\per\cm}) and WSe$\rm{_2}$ PL response (772 nm) with respect to the tip-sample distance, derived from (c), with fit to rate equation model as described in Supplementary Information.   
} 
	\label{fig:setup}
\end{figure}
%

\noindent As shown schematically in Fig. \ref{fig:setup}a, the experiment is based on a confocal microscope setup, with top illumination of a shear-force AFM tip for combined TERS and TEPL (see Methods and Fig. S1a for detail). 
A Helium-Neon laser beam (632.8~nm, $<$ 0.5~mW), after passing through a half wave plate for polarization control, is focused onto the WSe$\rm{_2}$ sample by an objective lens (100$\times$, NA=0.8).
The electrochemically etched Au-tip with typical apex radius (r $\sim15$~nm) is then positioned in the focal area \cite{neacsu2005}.
With the tip tilted by ${\sim}40$$^{\circ}$ with respect to the surface normal, confocal far-field or TERS and TEPL imaging and spectroscopy can then be performed alternatively, by simply retracting or engaging the plasmonic Au-tip with the sample.
The resulting localized surface plasmon resonance (LSPR) excitation in the axial detection of the tip with \textit{in-plane} sample projection of the locally enhanced near-field leads to effective excitation of the \textit{in-plane} Raman and exciton modes as characteristic for layered 2D materials, in contrast to conventional surface normal excitation for normal tip orientation (see Fig. S1b for details).
Fig. \ref{fig:setup}b shows the resulting anisotropy of the tip scattered plasmon response (blue) and far-field WSe$\rm{_2}$ PL (red) as a function of excitation polarization (black, slightly asymmetric due to the polarization-dependent incident optics) exhibiting the expected optical antenna behavior with excitation polarization parallel with respect to the tip axis (Experimental setup and observed spectra are shown in Fig. S1b).

Fig. \ref{fig:setup}c shows the distance dependence of TEPL and TERS of the ML WSe$\rm{_2}$ with representative PL spectrum (black) acquired at 6~nm distance, with Voigt profile fit dominated by the exciton response (${\sim}1.61$~eV) and a longer wavelength shoulder (${\sim}1.55$~eV).
The TEPL spectra reveal a slight redshift due to the softening of the coupled plasmon response \cite{rechberger2003}.
The shoulder at ${\sim}1.55$~eV is possibly due to biexciton (XX) emission \cite{you2015}. 
No spectral signature from the trion is seen or expected to be prominent at room temperature \cite{yan2014}. 
The near-field localization of both the TEPL and TERS response is shown in Fig. \ref{fig:setup}d for representative distances. 
The most prominent Raman peaks observed correspond to a superposition of E$^{'}$ (in-plane) and A$\rm{_{1}^{'}}$ (out-of-plane) modes at ${\sim}273$~\si{\per\cm} \cite{chen2015}, and first- and third-order LA phonons (M point in the Brillouin zone) at ${\sim}150$~\si{\per\cm} and ${\sim}405$~\si{\per\cm}, respectively \cite{zhao2013}.

Following the initially continuous increase in both TEPL and TERS response for $d\lesssim 20$~nm, at distances $d<5$~nm, the WSe$\rm{_2}$ PL starts to quench.
This behavior is due to a near-field polarization transfer between the WSe$_2$ exciton and the metal tip, giving rise to non-radiative damping and PL quenching. 
The PL distance dependence observed is well described by a rate equation model with damping rates $\Gamma$ and quantum yields of radiative emission $\eta$ corresponding to $1/\Gamma_{\mathrm{s}} \sim 0.5$~ps and $\eta_{\mathrm{s}} \sim 0.1$ for the sample, and $1/\Gamma_{\mathrm{tip}} \sim 30$~fs and $\eta_{\mathrm{tip}} \sim 0.5$ for the tip, and with the resonance energy transfer length of $R_0 \sim 8$~nm.
For TERS, the same set of parameters are used except assuming very short ($1/\Gamma_{\mathrm{s}} \sim 5$~fs) lifetime of the excitation to describe the instantaneous character of the Raman process (see Supplementary Information for details) \cite{kravtsov2014}.

\section{Results}
\begin{figure*}
	\includegraphics[width = 16cm]{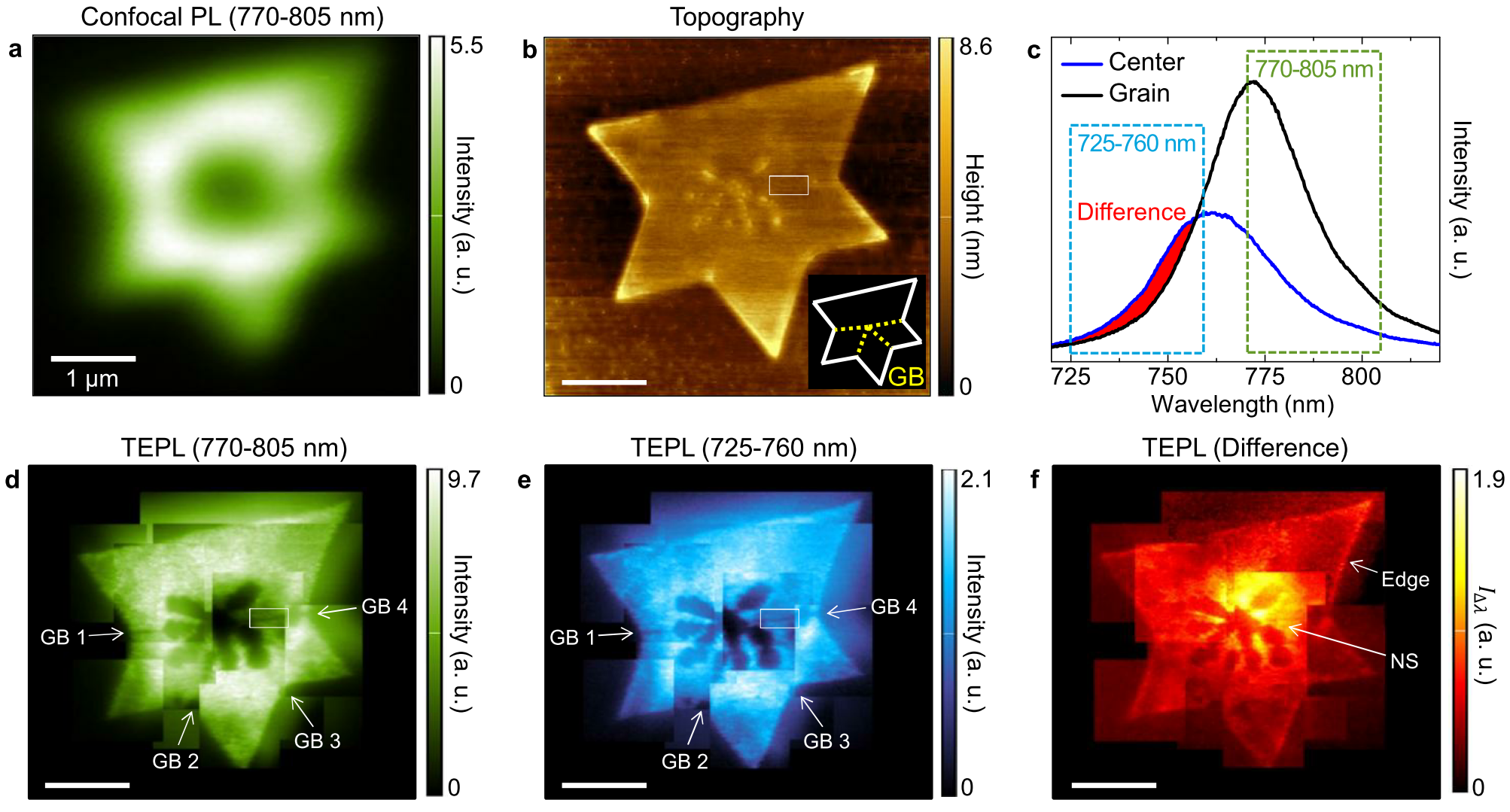}
	\caption{
(a) Confocal PL images of a ML WSe$\rm{_2}$ crystal for the integrated intensity at 770-805~nm (Blueshifted PL (725-760 nm) and spectral difference PL images are shown in Fig. S3a and b).
(b) Corresponding topography with inset illustrating the crystal and internal GBs.
(c) TEPL spectra of the ML WSe$\rm{_2}$ for center and grain regions. 
TEPL images for the integrated intensity of 770-805~nm (d) and 725-760~nm (e) spectral regions, and spectral difference TEPL image (f).
}
	\label{fig:tepl}
\end{figure*}

\subsection{Multi-Modal TEPL/TERS Imaging of Nanoscale Defects}
\noindent We then image NS, external crystal edges, and internal grain boundaries (GBs) through their effect on the PL and Raman response. 
Fig.~\ref{fig:tepl}a shows a ${\sim}500$ nm spatial resolution confocal PL survey of a polycrystalline ML
WSe$\rm{_2}$ flake with spectrally integrated 770-805~nm acquisition.
For the corresponding blueshifted PL image (725-760~nm), and difference PL image see Figs. S3b and c. 
Corresponding TEPL images with ${\sim}15$ nm spatial resolution reveal the influence of the NS region and crystal edge, which are seen to give rise to an associated decrease in PL intensity and spectral blueshift as shown in Fig.~\ref{fig:tepl}d-f (770-805~nm region, 725-760~nm region, and spectral difference image $\textit{I}_{\Delta \lambda}$).
Where $\textit{I}_{\Delta \lambda}=\int_{\lambda_1}^{\lambda_2} |{\textit{I}(\lambda)_{\textrm{i}}-\textit{I}(\lambda)_{\textrm{ii}}}| d\lambda$, $\lambda_1$ = 725~nm, $\lambda_2$ = 760~nm, and $\textit{I}(\lambda)_{\textrm{i}}$ and $\textit{I}(\lambda)_{\textrm{ii}}$ are the PL spectra for blueshifted center, edges and grains, respectively.

\begin{figure*}
	\includegraphics[width = 16cm]{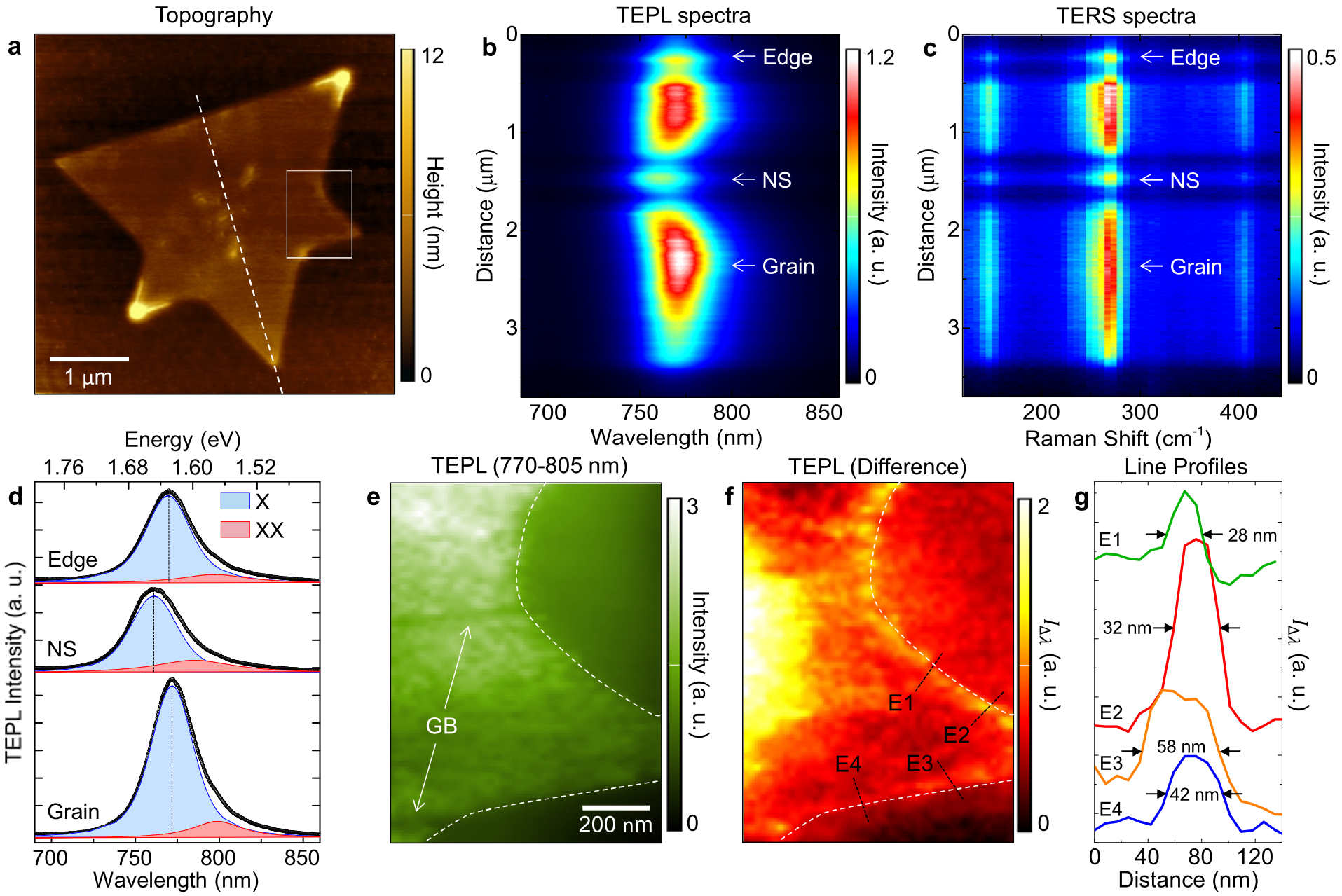}
	\caption{
Topography (a) of a ML WSe$\rm{_2}$ with spectral TEPL (b) and TERS (c)  line traces along the center line of the crystal (dashed line in (a)).
(d) Distinct TEPL spectra acquired from the edge, nucleation sites (NS), and grain region of crystal.
Exciton and possible biexciton peaks are assigned via the Voigt fitting. 
Peak energies are 1.610 eV at edge, 1.630 eV at center, and 1.606 eV at grain.
TEPL images of the spectral region of 770-805~nm (e) and the spectral difference (f). 
(g) Selected line profiles at the edge derived from (f). 
}
	\label{fig:spec}
\end{figure*}

Details of the effect of NS and crystal edges on the PL and Raman characteristics are investigated as shown in Fig.~\ref{fig:spec}, revealing in spatially-resolved spectral line traces (b, c) the decrease in PL and blueshift, being more pronounced for the NS compared to the edges (Fig.~\ref{fig:spec}d).
Intensity variations aside, no change in the corresponding Raman frequency of the E$^{'}$  and A$\rm{_{1}^{'}}$ superposition mode is observed (Fig.~\ref{fig:spec}c). 
The spatial variation of the PL at the edges (Fig.~\ref{fig:spec}e) is best exemplified in the spectral difference map (Fig.~\ref{fig:spec}f), with the PL shift and decrease extending irregularly along the edges, over a ${{\sim}{30-80}}$~nm wide region, as also seen in representative line traces E1-E4 (Fig.~\ref{fig:spec}g). 

\begin{figure}
	\includegraphics[width = 10cm]{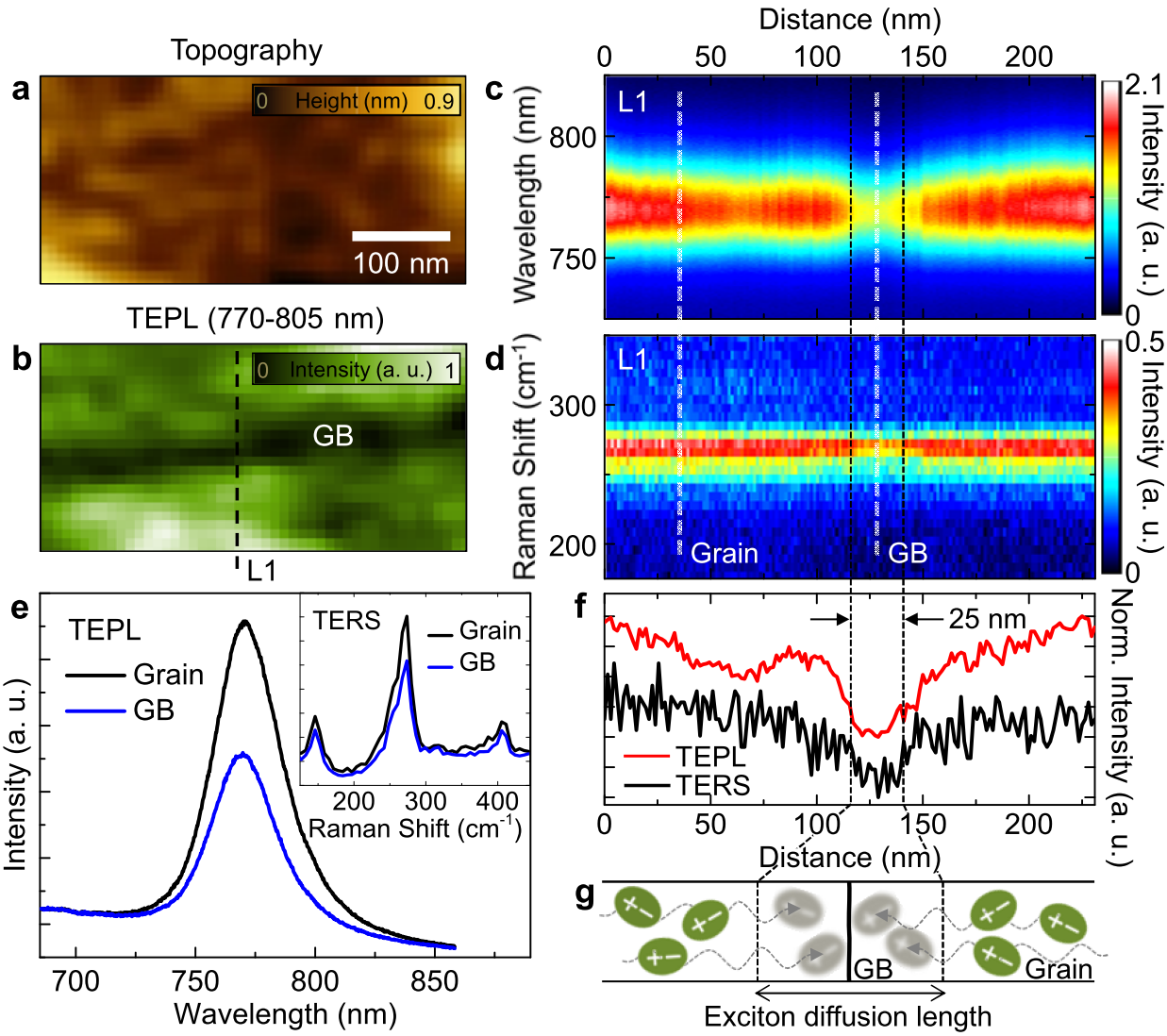}
	\caption{
(a) AFM topography of GB area (indicated in Fig. \ref{fig:tepl}b).
(b) TEPL image for the integrated intensity at 770-805~nm (indicated in Fig. \ref{fig:tepl}d).  
Spectral TEPL (c) and TERS (d) trace across GB (L1 indicated in (b)). 
(e) TEPL and TERS spectra derived from grain and GB regions indicated in (c) and (d) (dashed white lines).
(f) Corresponding spectrally integrated line profiles for the 770~nm PL peak and 273~\si{\per\cm} Raman mode derived from (c) and (d).
(g) Illustration of exciton diffusion length at GB. 
	}
	\label{fig:edge}
\end{figure}

The corresponding effect of the decrease in PL at the GBs can be clearly seen in the TEPL images Fig.~\ref{fig:tepl} and Fig.~\ref{fig:spec}e and appears somewhat distinct from the effect of edges and NS, as further analyzed in Fig.~\ref{fig:edge}, showing a zoom in to a single GB (white box in Fig.~\ref{fig:tepl}d).
The GB is not discerned in the AFM topography (Fig.~\ref{fig:edge}a).  
Neither PL nor Raman emission exhibit a spectral shift (Fig.~\ref{fig:edge}c-e), however both experience a decrease comparable to the case of NS and edges.
Yet the PL decrease is quite homogeneous along the grain boundary with a narrow width of ${\sim}25$ nm (Fig.~\ref{fig:edge}f). 

\subsection{Local Strain Engineering via Nano-Mechanical Force}

\begin{figure}
	\includegraphics[width = 16cm]{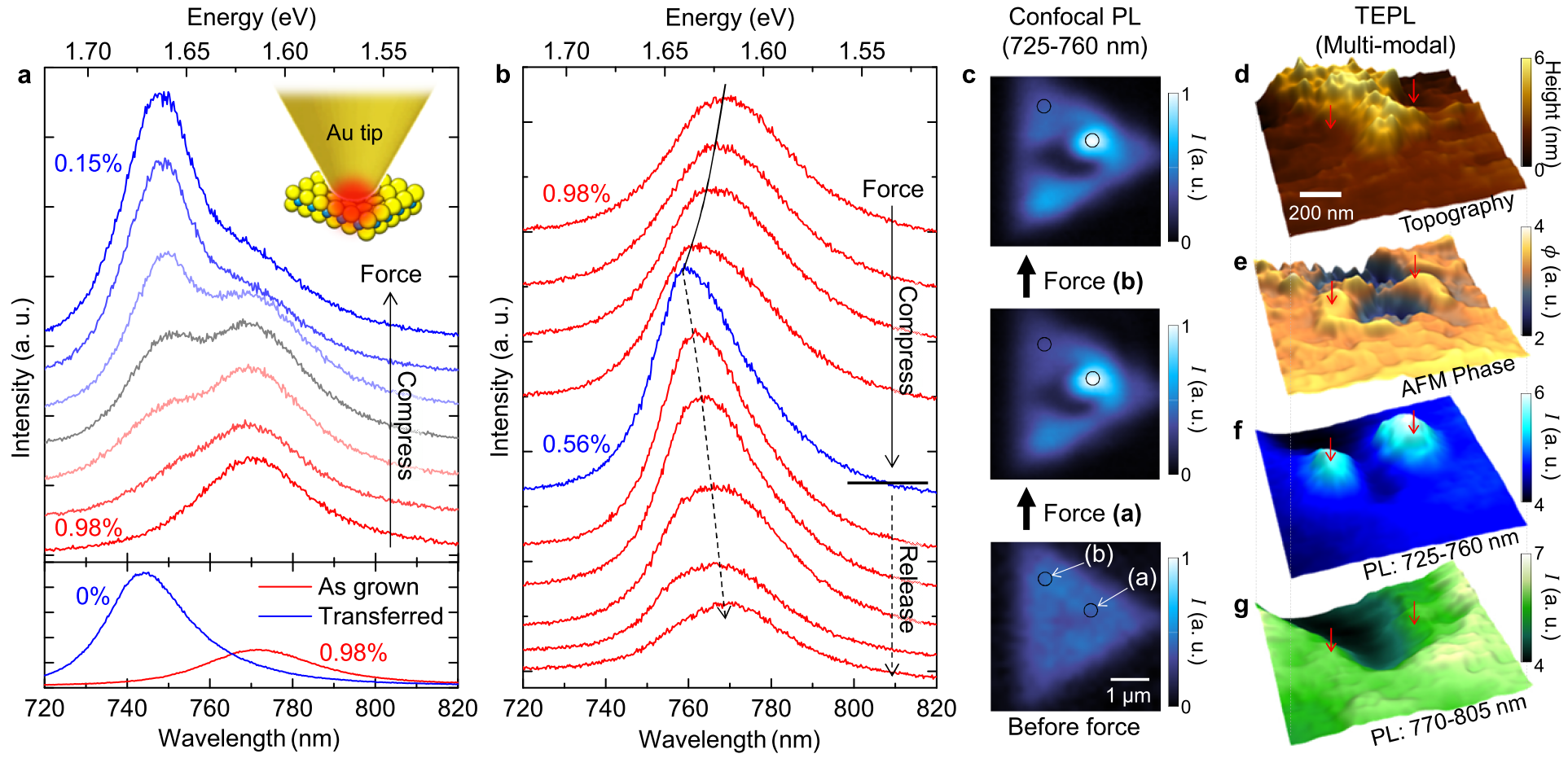}
	\caption{
	(a) Bottom: far-field PL spectra for the as-grown (red) and transferred (blue) ML WSe$\rm{_2}$ on the SiO$\rm{_2}$ substrates. 
	Top: evolution of PL spectra of the as-grown ML WSe$\rm{_2}$ with interacting compressive force of tip, giving rise to an irreversible release of the tensile strain of the crystal. 
	(b) Reversible evolution of PL spectra under modest nano-mechanical tip-sample force interaction. 
	(c) Confocal PL (725-760~nm) images before and after irreversible (a) and reversible (b) strain manipulation. 
	(d) AFM topography, (e) AFM phase, (f) blueshifted TEPL (725-760~nm), and (g) main TEPL (770-805~nm) images for two locations near NS which exhibit low PL prior to nano-mechanical strain release.
	}
	\label{fig:strain}
\end{figure}

\noindent In order to investigate the effect of local strain on the PL modification, we use the AFM tip to locally apply a contact force to perturb the sample while simultaneously measuring the TEPL. 
Fig.~\ref{fig:strain}a shows that with increasing force exerted by the tip, the TEPL increases and spectral weight is transferred with the appearance of a blueshifted emission by ${\sim}48$ meV. 
From comparison with far-field PL of both as-grown and transferred WSe$_2$ we assess that the crystals are initially under compressive strain of ${\sim}0.98$ \%. 
Thus the nano-mechanical tip interaction gives rise to an almost complete and irreversible strain release.
The PL modification at the strain released crystal region is seen in the confocal PL imaging before and after the force interaction (Fig.~\ref{fig:strain}c).
Despite the only few 10's nm spatial localization of the apex force interaction, the strain released crystal region extends spatially over a ${\sim}1~\mu$m sized area.
We estimate the maximum force exerted to ${\sim}10$ nN, without giving rise to tip apex modifications, as verified by repeated TEPL and TERS measurements.

Under careful and weak force interaction the strain can be partially released in a reversible fashion as shown in Fig.~\ref{fig:strain}b, with a spectral shift of ${\sim}24$ meV corresponding to a maximum release to ${\sim}0.56$ \%.
In this case also no discernible permanent PL modification is seen in the confocal PL image (Fig.~\ref{fig:strain}c). 
Both the reversible and irreversible strain release induced by the tip force interaction are generally repeatable with similar results for different flakes. 

Fig.~\ref{fig:strain}d-g shows TEPL imaging on a WSe$_2$ flake near a NS region with initially strongly suppressed PL.
Force interaction at the two locations indicated shows the appearance of blueshifted and enhanced PL. 
The associated strain release also manifests itself in a change in the AFM phase signal, yet not in the topography as expected.

\section{Discussion}

\noindent Our results are generally consistent with the findings of recent studies that have addressed the localized modification of the PL by structural heterogeneities \cite{van2013, liu2014, bao2015, Lee2015}, yet add further spatial details and assignment to specific processes. 
An {\em increase} in PL intensity correlated with a spectral blueshift at GBs observed in \cite{liu2014} has been interpreted in terms of a local strain release and structural deformation, associated with the large TMD crystals studied ($>$ 10 $\mu$m) and the resulting high substrate strain accumulated in the growth process. 
Our correlated spectroscopic results and the ease of nano-mechanical both reversible and irreversible local strain release by slight force perturbation would be consistent with this interpretation.

In contrast, recent near-field PL imaging revealed a {\em decrease} at GBs for small crystallites ($<$ 5 $\mu$m) \cite{bao2015} over a ${\sim}120$ nm average width and without spectral shifts. 
This, together with our finding of a PL quenching (47\%) without changes in structure and bandgap at the internal twin boundaries as concluded from the combined TERS and TEPL imaging, suggests that neither strain nor doping are responsible for the PL quenching, but instead non-radiative recombination process possibly from mid-gap states of defects. 
Given the atomic-scale lateral extent of the GB, the spatial scale of PL quenching would be related to the exciton diffusion length. 
Our value for the PL quenching width of ${\sim}25$~nm would be in good agreement with the expected exciton diffusion length of ${\sim}24$ nm as estimated from the measured values of exciton diffusivity and lifetime \cite{van2013}. 

At the crystal edges our observed decrease in PL intensity, spectral blueshift, and spatial heterogeneity agree with previous studies \cite{bao2015}. 
Similar to disordered semiconductors \cite{laquai2009, kagan1996} the quenching could be due to energy funneling from higher to lower energy states in the inhomogeneous system. 
Our spatial extent of the disordered PL modification region of ${\sim}{30-80}$~nm at the edges is in good agreement with the edge roughness observed by TEM \cite{van2013}. 
Note that our length scale is significantly shorter than the ${\sim}300$~nm reported in a previous near-field PL study \cite{bao2015}. 
This, and the shorter length scale observed at the GBs discussed above, suggests that the earlier transmission near-field scanning optical microscopy (NSOM) work was still spatial resolution limited \cite{bao2015, Lee2015}. 
Lastly, the significant PL quenching and blueshift at NS can be attributed to the loss of Se \cite{liu2015}, and the surface adsorbates are understood as tungsten (W) compounds left in the growth process \cite{cong2014} that can act as non-radiative recombination sites. 

The electronic band structure of TMDs is sensitive to tensile strain that causes reduced orbital hybridization due to the weakened ionic bonds \cite{he2013}. 
This gives rise to a redshift of the bandgap energy and an associated decrease in PL intensity \cite{cong2014, conley2013, gomez2013, desai2014}. 
Therefore, we believe the PL energy of transferred WSe$\rm{_2}$ is significantly higher than the as-grown sample due to the released tensile strain \cite{liu2014}, as confirmed by our force-controlled nano-spectroscopy measurements.

\section{Conclusions} 
\noindent In summary, we have measured modifications of the electronic structure and optical properties of WSe$\rm{_2}$ at the nanoscale through high resolution ($<$ 15 nm) multi-modal TEPL and TERS imaging. 
A nonlocal PL modification at twin boundaries associated with a ${\sim}25$~nm exciton diffusion length, and ${\sim}{30-80}$~nm wide region of optical heterogeneity at edges is observed. 
Further, we have demonstrated dynamic tuning of the local bandgap of ML WSe$\rm{_2}$ by releasing and controlling local strain. 
Our hybrid opto-mechanical nano-probe technique can be used for tunable nano-electronic devices where the carrier mobility is controlled via strain engineering \cite{cong2014, conley2013}.
We expect this method to help in the design of novel nano-photonic/electronic TMD devices by enabling local bandgap engineering and \textit{in-situ} spectroscopy of 2D materials.

\bibliography{tmd_2} 

\section{Acknowledgements}
\noindent The authors would like to thank Ronald Ulbricht for many insightful discussions and advise with the experiments. 
Funding was provided by the U.S. Department of Energy, Office of Basic Sciences, Division of Material Sciences and Engineering, under Award No. DE-SC0008807.

\section{Author contributions}
\noindent M. B. R., O. K. and X. X. conceived the experiment.
G. C. prepared the samples. 
K.-D. P. and V. K. performed the measurements, calculations, and analyzed the data.
K.-D. P., O. K., and M. B. R. wrote the manuscript with contributions from X. X.
All authors discussed the results and commented on the manuscript.
M. B. R. supervised the project.

{\section{Methods}}
\noindent
{\bf Sample Preparation}
WSe$_2$ monolayers are grown by physical vapor transport using powdered WSe$_2$ as precursor material. 
Source material (30 mg) in an alumina crucible is placed in the hot zone of a 25.4 mm horizontal tube furnace, and an SiO$_2$ substrate is placed downstream in a cooler zone at the edge of the furnace ($750-850$~$^{\circ}$C). 
Before growth, the tube is evacuated to a base pressure of 0.13 mbar and purged several times with argon. 
The furnace is then heated to 970~$^{\circ}$C at a rate of 35~$^{\circ}$C/min and remains there for a duration of 5-10 min before cooling to room temperature naturally. 
A flow of 80 sccm argon and 20 sccm hydrogen is introduced as carrier gas during the 5-10 min growth period. Details can be found in Ref. \cite{clark2014}.

\noindent
{\bf TEPL/TERS Setup} Fig. \ref{fig:setup}a shows a schematic of TEPL and TERS setup. 
The sample is mounted to a piezoelectric transducer (PZT, Attocube) for xyz scanning. 
Au tips are electrochemically etched to fabricate ${\sim}10$~nm apex radius \cite{neacsu2006}. 
Etched Au tips are attached to a quartz tuning fork (resonance frequency = ${\sim} 32$~kHz). 
The electrically driven tuning fork is vibrated with its resonance frequency and the changing amplitude signal due to the shear-force is monitored for tip-sample distance control \cite{karrai1995}. 
Tip positioning is operated by a stepper motor (MX25, Mechonics), and shear-force feedback and scanning conditions are controlled by a digital AFM controller (R9, RHK Technology).  
The incident laser beam is focused into the junction between the sample and the tip apex. 
The tip-enhanced PL and Raman signals are collected in backscattered direction, passed through an edge filter (633~nm cut-off) and detected using a spectrometer (f = 500 mm, SpectraPro 500i, Princeton Instruments) with a $N_{\rm2}(l)$ cooled charge-coupled device (CCD, Spec-10 LN/100BR, Princeton Instruments). 
The spectrometer is calibrated using hydrogen and mercury lines, and a 150 grooves/mm grating is used to obtain a high bandwidth spectrum for simultaneous measurement of TEPL and TERS.

\noindent
{\bf Multi-Modal Imaging} The PL peak of ML WSe$\rm{_2}$ has ${\sim}1.61$~eV bandgap (${\sim}770$~nm) and the PL intensity is uniform in the grain region. 
While, the PL intensity is decreased in the center and edge region, and the peak position is blueshifted (see Fig. S3a). 
To visualize the spatial heterogeneity of the PL spectrum, we propose the multi-modal PL imaging method. 
The integrated intensities for the main PL (770-805~nm) and the blueshifted PL (725-760~nm) are selectively counted at each pixel of sample scanning. 
However, the spatial heterogeneity of the blueshifted PL is not clearly visualized because the tail of the main PL still significantly affects the integrated intensity of the blueshifted PL. 
To solve this problem, we subtract the main PL image from the blueshifted PL image after compensating for intensity discrepancy. 
Through this method, the blueshifted PL property is clearly visualized in the PL images.

\noindent
{\bf Local Strain Engineering} To control the local strain of as-grown WSe$\rm{_2}$, we used the AFM tip of TEPL/TERS setup. 
The tip-sample distance is regulated by controlling the set-point and proportional-integral (PI) gains in feedback \cite{karrai1995}. 
We set a low set-point and a high PI gain to apply mechanical force to the ML WSe$\rm{_2}$. 
In this unusual feedback condition, AFM tip taps the sample with ${\sim}{10-30}$~Hz frequency and the tapping force is increased by lowering the set-point. 
Through this force control of tip, the strain of the as-grown WSe$\rm{_2}$ is locally released, as shown in Fig.~\ref{fig:strain}.

\end{document}